\documentclass[twocolumn,prl,aps,groupedaddress,nopacs,footinbib,preprintnumbers,floats,floatfix]{revtex4}

\usepackage{amsmath}
\usepackage{amssymb}
\usepackage{latexsym}
\usepackage{graphicx}
\usepackage{hyperref}
\usepackage{mathrsfs}
\usepackage{bm}

\begin{document}

\title{
A general test of the Copernican Principle
}

\author{ Chris Clarkson$^1$, Bruce Bassett$^{1,2}$, 
and Teresa Hui-Ching Lu$^1$\\
\it $^1$Cosmology \& Gravity Group, Department Mathematics
and Applied Mathematics, University of Cape Town, Rondebosch 7701,
South Africa.\\ 
$^2$ South African Astronomical Observatory, Observatory, Cape Town, South Africa}

\begin{abstract}
The recent discovery of apparent cosmic acceleration has highlighted the depth of our ignorance of the fundamental properties of nature. 
It is commonly assumed that the explanation for acceleration must come from a new form of energy dominating the cosmos - dark energy - or a modification of Einstein's theory of Relativity.  It is often overlooked, however, that a currently viable alternative explanation of the data is radial inhomogeneity which alters the Hubble diagram without any acceleration. This explanation is often ignored for two reasons: radial inhomogeneity significantly complicates analysis and predictions, and so the full details have not been investigated; and it is a philosophically highly controversial idea, revoking as it does the long-held Copernican Principle. 
To date, there has not been a general way of determining the validity if the Copernican Principle -- that we live at a typical position in the universe -- significantly weakening the foundations of cosmology as a scientific endeavor~\cite{george}. Here we present an observational test for the Copernican assumption which can be automatically implemented while we search for dark energy in the coming decade.  Our test is entirely independent of any model for dark energy or theory of gravity and thereby represents a model-independent test of the Copernican Principle. 
\end{abstract}
\maketitle

\paragraph{Introduction}

The Copernican Principle is at the root of cosmology, having played a pivotal role from the beginning of the modern history of the subject. Einstein erroneously imposed the Copernican Principle in both space and time by demanding a static universe, forcing him to introduce the repulsive cosmological constant. It is ironic that this very same cosmological constant may be responsible for the apparent accelerated expansion of the universe, thereby mysteriously placing the beginning of life on earth disconcertingly near the special cosmic time when this acceleration began. The magnitude of the cosmological constant required to achieve this is inexplicably small given our current understanding of quantum physics. Given these problems, might it not be just as natural for us to be living at a special place in a universe with no cosmological constant or dark energy, despite the charged religious connotations it would imply? Although violation of
the Copernican Principle would not solve the cosmological constant problem, it would
allow explanation of the data without the need for any dark energy. Of course, the bedrock of the standard model relies on the reasonable assumption that we are not living at a special place.
But to place cosmology on a firm scientific basis it is ever more clear that this matter must be settled observationally without theoretical bias. 

Until now this meant trying to develop inhomogeneous and/or anisotropic models to the level of sophistication that they could be tested against observations~\cite{alexander}~-- an extremely laborious task given that the Einstein Field Equations are nonlinear coupled partial differential equations. The usual argument for ignoring these difficult models comes from Occam's razor: why bother since we have a simpler model which is working so well? While radial inhomogeneity can fit any Hubble diagram perfectly~\cite{ltb,mustapha,szek}, in general it requires the introduction of an infinite number of free parameters to do so, clearly disfavored from a Bayesian point of view. However, it is not clear that one- or two-parameter models of radial inhomogeneity would not give as good a fit as homogeneous dark energy models with the same number of degrees of freedom, which leaves us in the same position of arbitrary model building that we have at present in the standard paradigm which instead parametrises an otherwise arbitrary equation of state for the dark energy, $w(z)$. 

Instead what we need is a model-independent test to determine if the Copernican assumption is in fact violated. One elegant suggestion to falsify the CP uses observations of the CMB from inside our light cone~\cite{goodman,caldwell}, either by testing for deviations from a blackbody spectrum, or deviations from isotropy at distant clusters. These important methods are very difficult and require precise control of systematic effects. Furthermore, these tests have a fundamental limitation. 
Even if we were to determine isotropy of the CMB for \emph{all} observers in the universe it would still not be enough to prove homogeneity, unless we also proved that all fluid components of the cosmic energy budget are co-moving, barotropic perfect fluids~\cite{CB}. 

We present here a new test for the CP which relies on a consistency relation that exists within the homogeneous and isotropic Friedmann-Lema\^\i tre-Robertson-Walker (FLRW) models between luminosity or area distance and the Hubble rate, both as a function of redshift. We show that this is not satisfied for radially inhomogeneous models in general, providing just such a mechanism to test whether we live in an approximation to an FLRW universe or, instead, near the center of a spherically symmetric one.

\paragraph{Consistency Relation}

In FLRW models, the luminosity distance may be written as (in units where $c=1$) 
\begin{equation}\label{d_L}
d_{L}(z)=\frac{(1+z)}{H_0 \sqrt{-\Omega_k}}\sin{\left( 
\sqrt{-\Omega_k}\int_0^z{\mathrm{d}z'\frac{H_0}{H(z')}}\right)},
\end{equation}
where $\Omega_k$ is the curvature parameter today, and the expansion rate $H(z)$ is given by the Friedmann equation and takes the value $H_0=H(0)$ today. The area distance is defined using $d_L=(1+z)^2d_A$, and another distance measure we will find useful is $D=(1+z)d_A$. We may rearrange Eq.~(\ref{d_L}) to give an expression for the curvature parameter in terms of $H(z)$ and $D(z)$:
\begin{equation}\label{OK}
\Omega_k=\frac{\left[H(z)D'(z)\right]^2-1}{[H_0D(z)]^2},
\end{equation}
where $'=\mathrm{d}/\mathrm{d} z$.
This tells us how to measure the curvature, in principle, from distance and Hubble rate observations, independently of any other model parameters or dark energy model.
Remarkably this tells us the curvature today from these measurements at any single redshift. 

Given that the curvature parameter is independent of redshift, we may differentiate this to obtain an expression which must equal zero. The factor responsible for this is 
\begin{equation}
\mathscr{C}(z)=1+H^2\left(DD''-D'^2\right)+HH'DD'\ ,
\end{equation}
which must be zero in any FLRW model at all redshifts, by virtue of Eq.~(\ref{d_L}). 
This function may also be derived by equating the two reconstructed  functions for $w(z)$ given by Eqs. (2) and (3) in \cite{CCB}.

As we have not utilised the Friedmann equation,  the derivation of $\mathscr{C}(z)$ relies only on the metric of spacetime, and not on the theory of gravity, nor on any matter model present~-- it is therefore a purely geometric function. Consequently, if the FLRW models are indeed the correct background model, then we should expect to measure $\mathscr{C}(z)\approx0$ (up to the amplitude of perturbations) in the real universe at all redshifts.

We may estimate the errors on $\mathscr{C}(z)$ using series expansions for $D(z)$ and $H(z)$, but with different FLRW parameter values, and dark energy equations of state. At leading order we have  
\[
\mathscr{C}(z)=\left[q_0^{(D)}-q_0^{(H)}\right]z+\mathcal{O}(z^2),
\] 
i.e., the difference between the separate deceleration parameters measured using $D$ and $H$. Hence, at low redshift, our error on $\mathscr{C}(z)$ is roughly the same as $q_0$ as estimated using $H(z)$. 

A critical aspect of our test is that $H(z)$ measurements must come from a different data set than used for distance data. Presently most estimates of $H(z)$ rely implicitly on Eq.~(\ref{OK}) or some other aspect of the FLRW models themselves. However, passively evolving objects such as luminous red galaxies provide a family of roughly synchronised clocks at moderate to high redshift, from which we can measure $H(z)$ directly using $H(z)=-1/(1+z)t'(z)$~\cite{H}. Alternatively, measuring the change of redshift, $\dot z(z)$, of an object over a period of several years can also provide $H(z)$~\cite{JP}, and so be used as a probe of the Copernican Principle based on the arguments presented here~\cite{UCE}. 

 Prospects for constraining $\mathscr{C}(z)$ in the near future as an independent confirmation of the standard model are exciting. For example, finding $w(z)$ from distance data requires knowledge of $D''(z)$, while reconstructing it from $H(z)$ alone requires knowing $H'(z)$~\cite{CCB}. Taking derivatives of observable functions introduces extra errors, so it might in practice be simpler to measure $\{\left[H(z)D'(z)\right]^2-1\}/{[H_0D(z)]^2}$ at different redshifts,  and check that it yields the same value as it should by virtue of Eq.~(\ref{OK}). In FLRW models this tells us the value of the curvature today; in more general models this quantity will not be as simple as this. This is an important consistency check on the FLRW models. If it is found to change with redshift then that is equivalent to finding  $\mathscr{C}(z)\neq0$~\footnote{We thank Ruth Durrer for pointing this out to us. Note that we must use the relation given by Eq.~(\ref{OK}) for this to work; measuring curvature at different redshifts by standard tests will be contaminated by our assumptions about the dark energy equation of state.}.  Thus, while we try to search for $w(z)$, we can measure $\mathscr{C}(z)$ at the same time to a similar degree of accuracy. 

But what would a measurement $\mathscr{C}(z)\neq0$ imply, even if only at one redshift? Since it's not dark energy or deviations from general relativity, the origin must be a different cosmological geometry which is not homogeneous. Clearly, then, measuring $\mathscr{C}(z)\neq0$ at any redshift would pose serious problems for cosmology.

We have given a necessary condition for violation of the Copernican Principle, but is it sufficient? In other words, could the principle be violated while still having $\mathscr{C}(z) = 0$? 
Consider Lema\^\i tre-Tolman-Bondi (LTB) models, which can fit the Hubble diagram without invoking dark energy~\cite{ltb}. These are spherically symmetric models with each shell at radius $r$ evolving as a separate FLRW model. Clearly the generic scenario is for each shell to be characterised by a different 
$\Omega_k$ and so have $\mathscr{C}(z) \neq 0$. Indeed, we show in the appendix that  $\mathscr{C}(z)$ is a freely specifiable function in these models, so $\mathscr{C}(z) = 0$ which are not FLRW is a very limited subclass within the LTB family.

\paragraph{Alternative Tests}

In addition to the key test just presented we present two other novel approaches to the problem.  
Future cosmological data will allow us to
measure $w(z)$ (assuming FLRW and an appropriate parametrisation) to
high accuracy \cite{detf}. With $w(z)$ known, one can make predictions for what will be observed with the
Alcock-Paczynski (AP) test \cite{AP}. The AP test relies on the idea
that a spherical object in real space will appear oblate or prolate in
redshift space due to the fact that the radial size, $L_{\|}$ is
determined by $\Delta z/H(z)$ while the transverse size, $L_{\perp}$ is
determined by $d_A(z)$. By demanding that $L_{\perp} = L_{\|}$ one can
determine cosmic parameters, in particular, $w(z)$.

The canonical AP test in modern cosmology is provided by the Baryon
Acoustic Oscillations (BAO) which uses the excess signal at $\sim
150$Mpc in the two-point correlation function as a standard ruler. By
observing this in both the angular and radial directions BAO will provide
both $d_A(z)$ and $H(z)$, and hence allow us to measure $w(z)$ to
similar accuracy as SNIa \cite{se}.

If the Copernican Principle is violated through significant radial
inhomogeneity then clustering in real space will not be isotropic. 
Hence the assumption underlying the AP and BAO tests (that the ``bump"
in the two-point correlation function is isotropic) will be invalid.
If one incorrectly assumes isotropic clustering one will reconstruct a
$w(z)$ that disagrees with that from SNIa or other measurements. Put another way
this would appear to violate distance duality (the equivalence between luminosity 
and angular-diameter distances) \cite{BK1}.  Tantalisingly such a mismatch has 
actually been observed between current BAO and SNIa data at about the $2\sigma$ level 
\cite{percival}. 

Indeed there are at least three ways in which violation of the Copernican Principle 
would deform standard BAO results. First, the sound horizon -- the fundamental standard ruler of the BAO method -- would be different in the $\|$ and $\perp$ directions (relative to the observer) and in general will differ from the standard FLRW value ($\sim150$Mpc today). Both of these will cause problems.  Secondly, the subsequent expansion of the sound horizon in the $\|$ and $\perp$ directions will be governed by different Hubble rates $H_{\|}$ and $H_{\perp}$ (see eq. \ref{H}). Hence, even if the sound horizon were isotropic and equal to the standard FLRW value at decoupling, the subsequent evolution would cause biases in measured cosmology. Finally, redshift distortions which need to be subtracted \cite{rd} in order to apply the AP test, will be significantly more complicated to model and crucially will be correlated with the background expansion, unlike the FLRW case. To see this consider a galaxy moving in a gravitational potential, $\Phi = \Phi_{\|} + \delta\Phi$ which is the sum of the usual perturbative gravitational potential and the radial inhomogeneity of the LTB background. Since $\nabla_{\perp} \Phi_{\|} = 0$, the $\|$ velocity of the galaxy will be different from the FLRW case even if the $\delta\Phi$ contribution is identical in both cases. This will modify the Doppler shift contribution to the redshift of the galaxy and hence will modify the usual Kaiser and ``Fingers of God'' effects in a way that is strongly correlated with the precise nature of the radial inhomogeneity.

Each of these effects will cause a bias in general in the estimation of dark energy parameters obtained by assuming a FLRW background which will manifest as a mismatch with the parameters derived from distance measurements derived from supernovae or lensing. It is also perhaps worth noting that the growth rates will be different in the $\|$ and $\perp$ directions.

A significantly more exotic test for the Copernican Principle is
provided by the realisation that the $d_L(z)$ relation need not be
single-valued; see Figs 5-8 of \cite{mustapha}. The same redshift can
correspond to more than one distance. Clearly in FLRW this is
impossible and the occurance is related to the possibility of
gravitational blueshifts/redshifts adding to the standard expansion
redshift. Hence, object A which is further away than object B, may
still have the same redshift if it lies at a suitably higher
``potential energy" which compensates the difference in cosmic
expansion. Hence, the volume as a function of redshift can exhibit pathological behaviour which would be visible in number-counts or a large dispersion in SNIa distances around the same redshift providing a clear signal of violations
of the CP.

\paragraph{Discussion}

On what timescale can we expect these various tests to be implemented? 
Future BAO surveys such as WFMOS will measure $d_A(z)$ and
$H(z)$ to better than $2\%$ and large radio surveys such as HSHS and
SKA will measure them to better than $1\%$ in redshift bins of width
$\Delta z \sim 0.2$ \cite{hshs,ska}. At its simplest our test can be implemented by measuring 
$\Omega_k$ at two different redshifts. Combinations of future all-sky lensing, BAO and SNIa 
surveys will allow a measurement of $\Omega_k$ with accuracy $\sim 0.04$\cite{bernstein} which
sets the approximate scale of how well we will test the Copernican Principle in the next decade and which 
should allow us to strongly rule out inhomogeneity as the source of oversized cosmic distances.  
Beyond these surveys one can envisage large, low-frequency HI surveys able to measure $d_A(z)$ and $H(z)$ at high redshift, 
$z \sim 10-100$ which will allow the testing of the Copernican Principle 
in the first billion years of cosmic history when most perturbation modes were still in the linear regime.

While the Copernican assumption is unlikely to be dramatically wrong in our Hubble sphere, it is possible that the dark energy phenomenon is actually concealing something even more bizarre than a misconception of fundamental physics. Indeed, with suggestions such as chaotic inflation, everlasting bouncing universes and the like, it is statistically possible that our part of the universe has some unusual features on a spatial, instead of a purely temporal, dimension, see e.g. ~\cite{linde}. This cannot be ruled out at the moment on anything other than philosophical grounds.  
 
We have presented a new, straightforward, test of the Copernican assumption which may play an important role in our understanding of dark energy.
While we strive to determine the function $w(z)$, we have shown that we can simultaneously constrain our Copernican function $\mathscr{C}(z)$ to a similar degree of accuracy. At any redshift, a measurement of $\mathscr{C}(z)\neq0$ would imply that the FLRW models are the wrong foundation for cosmology and that something more sophisticated must be considered instead. Even if the cosmos is FLRW on average, measurement of $\mathscr{C}(z)$ at small redshifts will allow us to probe the scale at which homogeneity sets in. The proposed test does not depend on any theory of gravity nor on our understanding of dark energy, but relies only on the geometry of the FLRW models themselves. 

\acknowledgements
We thank  Marina Cort\^es, Ruth Durrer, George Ellis, Pedro Ferreira, Ariel Goobar, Charles Hellaby, Roy Maartens, Bob Nichol and  Jean-Phillipe Uzan for comments and Charles Hellaby for providing the code used in \cite{LH}. This research is funded by the NRF (South Africa).

\appendix

\paragraph{Appendix: $\mathscr{C}(z)$ in LTB models}

The general spherically symmetric metric for an irrotational dust
matter source is the
Lema\^{\i}tre-Tolman-Bondi (LTB) metric, given in  synchronous comoving coordinates as
\[
ds^{2} = -dt^{2} +\frac{\ [\partial_r R(t,r)]^{2}}{1 + 2E(r)}\ dr^{2} +
R^{2}(t,r)d \Omega^{2}\ , \label{LTBmetric}
\]
where $\partial_r R(t,r) = \partial R(t,r)/\partial r$, and $d \Omega^2 = d
\theta^2 + \sin^2 \theta d \phi^2$ is the metric on a unit 2-sphere. The function $R
= R(t,r)$ is the areal radius. The function $E = E(r) \geq -1/2$ is an arbitrary function of the LTB model, representing the local curvature.

Solving the Einstein field equations, we have a generalized Friedmann equation for the \emph{angular} Hubble rate: 
\begin{equation}
H_\perp(t,r)=\frac{\partial_t R(t,r)}{R(t,r)} = {R(t,r)}^{-1}\sqrt{\frac{2M(r)}{R(t,r)}+2E(r)}\ , \label{Rdot}
\end{equation}
where $M(r)$ is another arbitrary function of the LTB model that
gives the gravitational mass within comoving radius $r$, which may be derived given the energy density $\rho(t,r)$. There is one further free function of radius, the bang time. Of the three radial degrees of freedom, one is gauge, representing our choice of radial coordinate, while two are genuine physical degrees of freedom, and can be specified arbitrarily. We shall assume that the coordinate degree of freedom removes (or specifies) the bang time function: thus, specifying $\{M(r),E(r)\}$ fully specifies the LTB model. 

In LTB models the expansion in the radial direction, $H_\|(t,r)$, is not the same as that in the angular direction, but is given instead by~\cite{MT}
\begin{equation}
H_\|(t,r)=\frac{\partial_t\partial_r R(t,r)}{\partial_rR(t,r)}
=H_\perp(t,r)+\frac{\partial_r H_\perp(t,r)}{\partial_r \ln R(t,r)}; \label{H}
\end{equation}
 in the FLRW limit $\partial_r  H_\perp(t,r)=0$, so the two Hubble rates are the same. This can present different ways to view $\mathscr{C}(z)$ in LTB models, as we can use whichever $H(z)$ we like. 

Finding the area distance explicitly in terms of redshift is not simple, and is discussed in detail in~\cite{mustapha}. In the usual procedure, one chooses a radial coordinate which flattens out the central observer's null cone. Evaluating the function $R$ on the null cone then turns it into the area-distance as a function of redshift: $d_A(z)=R|_{\mathrm{null~cone}}$. However, it has been shown in~\cite{CR} that $d_L(z)$, or equivalently $d_A(z)=R(z)$ (and so $D(z)$), may be considered as a free function of the LTB model instead of $M(r)$. Therefore, the LTB model may be directly specified by the free functions $\{D(z), E(r)\}$. On the null cone we may find the radial coordinate $r(z)$, implying that $E(r(z))=E(z)$ is a free function instead of $E(r)$. As a free function, we may replace $E(z)$ with $H_\perp(z)$ using Eq.~(\ref{Rdot}) evaluated on the null cone, which means that an LTB model can in fact be specified by the free functions $\{D(z),H_\perp(z)\}$ or $\{D(z),H_\|(z)\}$. 

So, as far as $\mathscr{C}(z)$ is concerned, we now see that in an arbitrary LTB model, it can be \emph{anything} one chooses, whether we are to interpret $H(z)$ as $H_\perp(z)$ or $H_\|(z)$, or a suitable combination of them.\footnote{We have verified that in fact $\mathscr{C}(z)$ is non-zero for a class of LTB models, using the code presented in~\cite{LH}.}
 If it is zero for both $H_\perp(z)$ and $H_\|(z)$, potentially giving a class of LTB models which would fail our Copernican test, then that leaves just one degree of freedom~-- exactly as in FLRW models with dark energy. If this free function is $H(z)=H_\perp(z)=H_\|(z)$, say, then $D(z)$ must be given by Eq~(1), which can be shown by integrating $\mathscr{C}(z)=0$. Although we have not shown that these models are necessarily FLRW, this would have to be a very restricted family within the full LTB class.  It would be interesting to determine the exact conditions under which $\mathscr{C}(z)=0$ observed from one location is a sufficient condition for an expanding spacetime to be FLRW. 

This `Copernican function' therefore must be considered, as far as a test for the CP is concerned, as essentially free, and must therefore be determined by observations.

\end{document}